\input amstex


\def\({\left(}
\def\ld{\left.}

\def\<{\left\langle}
\def\lv{\left|}
\def\){\right)}

\def\>{\right\rangle}
\def\rv{\right|}

\def\wt{\widetilde}
\def\ol{\overline}


\def\b1{\text{\bf 1}}

\def\dpar{\partial}

\def\#{\,\check{}}

\def\fC{{\frak C}}

\def\Ci{\fC_{it}}
\def\Cis{\overset{\smile}\to{\fC}_{it}}

\def\vphi{\varphi}
\def\veps{\varepsilon}
\def\wtC{\wt{C}}
\def\wtfC{\wt{\fC}}
\def\wtD{\wt{D}}
\def\wtM{\wt{M}}
\def\wtT{\wt{T}}
\def\wtQ{\wt{Q}}


\predefine\divs{\div}
\def\geqs{\geqslant}

\def\leqs{\leqslant}


\def\div{\operatorname{div}}
\def\pderby#1{{\dpar}\over{\dpar{#1}}}

\def\abs#1{\lv{#1}\rv}

\documentstyle{amsppt}

\NoBlackBoxes

\topmatter
\title     The quantum Cauchy functional and  space-time approach to relativistic quantum mechanics    \endtitle
\author A.~A.~Beilinson  \endauthor
\leftheadtext{A.~A.~Beilinson}

\affil Peoples' Friendship University of Russia \\ Department of Theoretical Physics an Mechanics \\ Moscow \\ {\it E-mail: alalbeyl{\@}gmail.com} \endaffil
\address Peoples' Friendship University of Russia, Moscow \endaddress
\email alalbeyl@gmail.com \endemail
\date \enddate
\thanks \endthanks
\translator \endtranslator
\subjclass 46T12, 81Q05 \endsubjclass


\keywords the momentum and coordinate presentations of functionals, Parseval's identity, generalized quantum Cauchy process, generalized pre-measure, 
generalized measure, generalized  countably additive measure, the support of a generalized measure, generalized path integral, Zitterbewegung, 
the Foldy-Wouthuysen transform
 \endkeywords

\abstract We construct generalized quantum Cauchy pre-measures that correspond to the analytic continuation of 
the transition probability of the Cauchy process to imaginary time.
We show that these complex pre-measures of time translations extend to a measure on the space dual to a real Hilbert space whose support is locally compact in the uniform convergence topology and with velocities in the Hilbert space.
 
 At that the quantum Cauchy-Dirac and Cauchy-Maxwell pre-measures of time translations of electrons and photons, that correspond to the retarded Green's functions of the Dirac and Maxwell equations with no sources viewed as generalized functions on bump functions, are unitary equivalent to quantum Cauchy pre-measures

Therefore these pre-measures  on the space dual to a real Hilbert space are $\sigma$-additive as well, but their support on the electron (respectively, photon) trajectories are compact in the uniform convergence topology with velocities in the Hilbert space.

 We study the way the classical relativistic mechanics of particle comes from the quantum mechanics of the free Dirac particle.   \endabstract

\endtopmatter

\document


\head  Introduction \endhead

We study one-and three-dimensional quantum Cauchy functionals, that are well-defined analytic continuations to imaginary time axis of the transition probability of the Cauchy process (see \cite{7}, \cite{15}), as   non-Gaussian  scalar
 complex-valued functionals on bump functions. 

We use the next important property of these functionals, which are  
fundamental solutions for  integral evolution equations, as well as measurable functions:  They generate compatible complex normed pre-measures of cylindrical subsets with Borel bases in $R^{(n)}$ (or in  $R^{(3n)}$ in the space case)  that are Lebesque's integrals over the bases.

These  comlpex normed pre-measures    corresponding to the
quantum Cauchy functional,
  are such that the measure of the complement to a ball of large enough radius $R$  in a Eucledean space  of any dimension, together with the measure of any its subdomain, is purely imaginary of fixed sign and tends to zero when $R\to \infty$.

Therefore to these complex pre-measures of translations there correspond the balls of finite radius in $L_2 (0,t)$ whose complements have measure smaller than any given $\veps >0$ and so are  weakly compact, and the pre-measures in $R^{(n)}$, despite being complex-valued,  can be extended to a space dual to a real Hilbert space. At that a set of time trajectories that has full measure is  locally compact in the uniform convergence topology  with derivatives in  $L_2 (0,t)$.

Though the 3-dimensional problems for  quantum Cauchy functionals are not the product of 1-dimensional ones, it is clear that all the results obtained for 1-dimensional quantum Cauchy functional, after proving the $\sigma$-additivity of the corresponding pre-measure, are automatically true for the space case. So
we will describe in details only the 1-dimensional case.

\medskip

At the end of the article, using an isomorphism between the spaces of solutions 
of the Dirac equation for free electron and the Maxwell photon equation and the respective presentations of Foldy-Wouthuysen (which are diagonalized solutions that can be reduced to quantum Cauchy functionals), we show that all the results render to these physical processes of relativistic quantum mechanics.

\medskip

At that we find that the support of the measures of trajectories of relativistic particles  restricts to the ball of radius $t$ in the space of continuous functions $C(0,t)$ while preserving the fundamental local property of the support of that measure - that the velocities lie in the Hilbert space.

Everywhere (except the last section) we use the system of units with the speed of light $c=1$ and Planck constant $\hbar =1$.

\medskip

For a preliminary exposition of the results see \cite{23}.

\head 1. The quantum Cauchy functionals and relativistic quantum mechanics of Dirac's electrons and Einstein's bosons. \endhead

One knows that the momentum presentation $\wtD^m_t (p)$ of the fundamental solution $D^m_t (x)$ of Dirac's  free electron equation

$$
i{\partial\over{\partial t}} D^m_t (x)=(i^{-1}\gamma^0 (\gamma ,\nabla )+\gamma^0 m) D^m_t (x) ,  \tag 1
$$
 where $\gamma^0 ,\gamma^1 ,
\gamma^2 , \gamma^3 =
\gamma^0 ,\gamma$ are Dirac matrices (see \cite{14}), can be written as 
$$
 \wtD^m_t (p)=\wtT^m (p)  \wtD^{m^F}_t (p) \wtT^m (p)   .\tag 2
$$
 
Here $ \wtD^{m^F}_t (p) =\exp(it\gamma^0 \sqrt{m^2 +\rho^2})$, $\rho =\abs{p}$, is the momentum Foldy-Wouthuysen presentation, and 
$ \wtT^m (p)=\wtT^{m^{-1}} (p)=\gamma^0 
 {{(p,\gamma )+ I(m+\sqrt{m^2 +\rho^2})}\over{\sqrt{2\sqrt{m^2 +\rho^2}(m+\sqrt{m^2 +\rho^2})}}}$ is a unitary operator.

If  we view $ \wtD^m_t (p)$ as a functional on bump functions $\varphi (x)\in K$ then one can construct the coordinate presentation $D^m_t (x)$
as an analytic functional on $\psi (x)\in Z$ (see \cite{2}) if we define  $C^m_{it}(x)$ as Fourier transform of $\wtC^m_{it}(x)=\exp (it\sqrt{m^2 +\rho^2})$ using the Parseval identity (see \cite{2}). Here the key role is served by the integral (see \cite{15})

$$
\int_0^\infty \exp (-t\sqrt{m^2 +\rho^2})  \cos (px) dp={{tm}\over{\sqrt{m^2 +x^2}} }K_1 (tm \sqrt{t^2 +x^2}) 
$$
 (here $K_1 (z)$ is the Macdonald function, see \cite{9}, 3.7 formula (6), and $t$ and $x$ are real, $t>0$) and the formula for the flat wave decomposition of $\delta$-function, see \cite{2}.

At that the constructed analytic functional $C^m_{it}(x)$ (and hence  $D^m_{it}(x)$ - see (2)) is defined on the bump functions if the singular integrals (due to the behavior of the Macdonald function on the light cone - $\ld{K_1 (z)}\rv_{z\to 0} \simeq z^{-1} $ - see \cite{10}) as regularized in the sense of Cauchy's principal value - see \cite{2}. Therefore one has 
$$
\int \overline{C^m_{it}(x)}\varphi (x) dx= \overline{\varphi (x)}^{S_t} +{1\over {\pi^2}} \int\overline{{{\partial}\over{\partial (l_t^2 )}}{{-tmK_1(iml_t )}\over{l_t}}}\varphi (x) dx  \tag 3
$$
 where  $\overline{\varphi (x)}^{S_t}$ means averaging of    $\varphi (x)\in K$  over the sphere of radius $t$, $l_t =\sqrt{t^2 - r^2}$, $r=\abs{x}$. Here the  momentum presentation $\wtC^m_{it} (p)$, as well as $\wtD^{m^F}_{t} (p)$, is an analytic functional on $Z$. Notice that at $m=0$ equality (3) becomes 
$$
\int \overline{C_{it}(x)}\varphi (x) dx= \overline{\varphi (x)}^{S_t} -{i\over {\pi^2}} \int {t\over{(t^2 - r^2 )^2}} \varphi (x) dx  \tag 4
$$
 and $\wtC_{it} (p)=\exp (it\rho )$.

Therefore $C_{it}(x)$ can be naturally viewed as well defined  analytic continuation to imaginary time
axis of   the transition probability ${1\over{\pi^2} }{t\over{(t^2 +r^2)^2}}$ of the space Cauchy process (see \cite{7}) as a functional on $K$. So we will call the generalized retarded Green's function 
$C_{it}(x)$ the quantum Cauchy functional, and call $C^m_{it}(x)$
the modified quantum Cauchy functional. 

Clearly $C_{it}(x)\neq \Pi^3_{j=1} C_{it}(x_j )$ and the support of that functional does not lie in Minkowski space.

Notice that in 1-dimensional case one has 
$$
C_{it}(x)={1\over 2} (\delta (t-x)+\delta (t+x))+ {i \over \pi} {t\over {t^2 -x^2}}, \quad \wtC_{it}(p)=\exp (it\abs{p}), \tag 5
$$

$$
C^m_{it}(x)={1\over 2} (\delta (t-x)+\delta (t+x))+ {1 \over \pi} {{-tm K_1 (im \sqrt{t^2 - x^2})}\over {\sqrt{t^2 -x^2}}}.
$$

The Maxwell equations 
$$
{\partial\over{\partial t}} E_t (x) =rot H_t (x),\quad {\partial\over{\partial t}} H_t (x) =-rot E_t (x) \tag 6
$$
 for the photon field in Majorana variables (see \cite{1}) $M_t (x)= E_t (x)+i H_t (x)$, $\ol{M}_t (x)= E_t (x)-i H_t (x)$ look in the momentum variables $\wtM_t (p)$, $\ol{\wtM}_t (p)$
as follows (see \cite{1}): 
$$
i{\partial\over{\partial t}}
\wtM_t (p)=(S,p)\wtM_t (p),\quad i{\partial\over{\partial t}}
\ol{\wtM}_t (p)=-(S,p)\ol{\wtM}_t (p).
$$
 Here $(S,p)=\Sigma_{j=1}^3 s^j p_j$ and 

$$
 s^1 = 
\pmatrix
0 &0&0 \\
 0&0&-i\\
 0&i& 0
\endpmatrix , s^2 = \pmatrix
0 &0&i \\
 0&0&0\\
 -i&0& 0
\endpmatrix , s^3 = \pmatrix
0 &-i&0 \\
 i&0&0\\
 0&0& 0
\endpmatrix 
$$
 are spin operators of the photon so $(S,p)$ is Hermitian.

The roots of the characteristic polynomial of $(S,p)$ are $\pm \rho$ ($\rho =\sqrt{p_1^2 + p_2^2 + p_3^2}$) and 0, so this matrix is degenerate. One has 
$$
(S,p)= \wtQ^+ (p)\tilde{h}^F (p)\wtQ(p)
$$
 where 
$$
\tilde{h}^F (p)= 
 \pmatrix
\rho &0&0 \\
 0&-\rho &0\\
 0&0& 0
\endpmatrix ,
$$
  $\wtQ (p)$ is a unitary operator that diagonalizes $(S,p)$, and $\wtQ^+ (p)$ is the adjoint operator.

\medskip

Therefore the momentum presentation of the fundamental solutions in Majorana variables is the direct product of matrices 
$\tilde{\text{M}}_t (p)=\wtM_t (p)\times \ol{\wtM}_t (p)$
where 
$$
 \wtM_t (p)= \wtQ^+ (p)\wtM_t^F (p)\wtQ(p), \quad \wtM_t^F (p)= 
 \pmatrix
\exp (-it \rho ) &0&0 \\
 0&\exp(it\rho )&0\\
 0&0& 1
\endpmatrix , \tag 7
$$

$$
\ol{\wtM}_t (p)= \wtM_{-t }(p),
$$
 and $\wtM_t^F (p)$ can be called the momentum Foldy-Wouthuysen presentation of the photon fundamental solution.

If $\tilde{\text{M}}_t (p)$ is understood as an analytic functional on $Z$ then  M$_t (x)$ is a functional on $K$ and one has 
$$
\text{M}_t (x)=  Q^+ (x) * M^F_t (x)* Q(x) \times  Q^+ (x) * \ol{M}^F_t (x)* Q(x)
$$
 where the functional (to be compared with $D_t^{m^F}(x)$)  
$$
M^F_t (x)=\pmatrix
\ol{C}_{it}(x)  &0&0 \\
 0&C_{it}(x) &0\\
 0&0& \delta (x)
\endpmatrix
$$
 has support that does not lie in the Minkowski space.

\medskip

Therefore  the study of matrix-valued fundamental solutions of   relativistic quantum mechanics equations for the electron and photon is reduced to one of scalar quantum Cauchy functionals.

\head 2. The quantum Cauchy functional; the constructions of pre-measures   \endhead

We study the position presentation (5) of the one-dimensional quantum Cauchy functional $C_{it}(x)$ on the bump functions $\varphi (x)\in K$ (here the second summand is understood to be regularized, see \cite{2}), its momentum presentation $\wtC_{it}(p)=\exp(it\abs{p})$ being a functional on analytic test functions $\psi (p)\in Z$.

Our aim is to construct and study first complex-valued measure of time translations of a particle (cylindrical  sets of trajectories) that correspond to 
$C_{it}(x)$ with $n$-dimensional Borelian bases (pre-measures) whose points $a_k$, $k=1,\ldots ,n$, are given by conditions 
$$
 a_k= \sum_{1\le j\le n} \alpha_{jk}\Delta x_j, \quad k\le l\le n, \det \alpha_{jk}\neq 0 ,\tag 9
$$
 where $ \alpha_{jk} $ are real numbers and $ \Delta x_j =x_j - x_{j-1}$ are changes of the position of the particle (the translation) for the time $ \Delta t_j =t_j - t_{j-1}$ ($j=1,\ldots ,n$, $ t_0 =0, t_n =t$) or $a_k =\int_0^t 
\alpha_k (\tau )dx (\tau )$, $
\alpha_k (\tau )$ are piecewise constant functions.

We first construct   retarded Green's functions $\fC_{it}(a_1 ,\ldots , a_n )$ that correspond to $C_{it}(x)$ (see (5)) as functionals on $  K $ on $n$-dimensional base of the set of translations  (11). We will see that  these Green's functions (as measurable functions) define measures of the   cylindrical sets of trajectories with Borel bases  in $R^{(n)}$.

Recall that $C_{it}(x)$ on $  K $ is a retarded Green's function (see \cite{5}) hence satisfy the Chapman-Kolmogorov equation 
$$
 \int \overline{C_{i\Delta{t_1}}(\Delta x_1 )\ldots C_{i\Delta{t_n}}(\Delta x_n )}\varphi (\Delta x_1 +\ldots +\Delta x_n ) d\Delta x_1 \ldots d\Delta x_n =\int 
\overline{C_{it}(x)}\varphi (x)dx , \tag 10
$$
 (i.e., $C_{i\Delta{t_1}} *\ldots *C_{i\Delta{t_n}} = C_{it}$ where $*$ is the convolution, see \cite{2}) where $0=t_0 < t_1<\ldots <t_n =t$, $\Delta t_j = t_j - t_{j-1}$, $j=1,\ldots , n$. Since in the r.h.s.~of (10) one has $x=\Delta x_1 +\ldots +\Delta x_n$, the argument of $C_{it}(x)$ is a point on the base of of the simples cylindrical subset of trajectories $\Sigma^n_{j=1} \Delta x_j =x$.

\medskip

So to construct the retarded Green's function $\fC_{it}(a)$ on one-dimensional base in the general case that corresponds to  $\Sigma^n_{j=1} \alpha_j \Delta x_j =a$, one   considers 
$$
\int{\(\prod\limits_{j = 1}^{n}{\ol{C_{i\Delta{t_j}}(\Delta{x_j})}}\)\vphi\(\sum\limits_{j = 1}^{n}{\alpha_{j}\Delta{x_j}}\)d\Delta{x_1} \ldots d\Delta{x_n}}\text{.}
\tag 11
$$

Since $\vphi\(\sum\limits_{j = 1}^{n}{\alpha_{j}\Delta{x_j}}\) = \frac{1}{2\pi}\int{\exp{\(-ip\sum\limits_{j = 1}^{n}{\alpha_{j}\Delta{x_j}}\)}\psi(p)dp}$, (11) implies 
$$
\int{\(\prod\limits_{j = 1}^{n}{\ol{\wtC_{i\Delta{t_j}}(p\alpha_{j})}}\)\psi(p)dp} = \int{\ol{\wtC_{i\sum\limits_{j = 1}^{n}{\abs{\alpha_{j}}\Delta{t_j}}}(p)}\psi(p)dp} \text{,}
$$
 hence 
$$
\fC_{it}(a) = C_{iT}(a) \text{,\quad} T = \sum\limits_{j = 1}^{n}{\abs{\alpha_{j}}\Delta{t_j}} \text{.}
\tag 12
$$
 Here the
retarded Green's function $\fC_{it}(a)$ is a functional on the space $K$ of bump functions.

\medskip

{\it Remark.} As was already noticed, the retarded Green's function $C_{it}(x)$, viewed as a functional on $\varphi (x)\in K^{(1)}$, is assumed to be regularized, so $\int ((t-x)^{-1}+(t+x)^{-1} )\varphi (x)dx$ is understood in the sense of the Cauchy principal value (see \cite{2}). So it is convenient to present  $C_{it}(x)$ as 
$$
\gathered
   C_{it}(x) = \frac{1}{2}\(C_{it}^{-}(x) + C_{it}^{+}(x)\) \text{,} \\
   C_{it}^{\pm}(x) = \delta(t \pm x) + \frac{i}{\pi}\frac{1}{t \pm x} = \cases
                                                                           \delta(t \pm x) \text{,\quad} \abs{t \pm x} < \veps \\
                                                                           \frac{i}{\pi}\frac{1}{t \pm x} \text{,\quad} \abs{t \pm x} \geqs \veps \text{,\ } 0 < \veps \to 0 \\
                                                                           \frac{i}{\pi}\frac{1}{t \pm x} \text{,\quad} \abs{x} < A \text{,\ } 0 < A \to \infty \\
                                                                           0 \text{,\quad} \abs{x} \geqs A
                                                                        \endcases \text{.}
\endgathered
\tag 13
$$
 Thus the integral of $C_{it}(x)$ over every $\varepsilon$-interval is well defined.
Therefore $C_{it}(x)$ yields complex measure $\int_R \fC_{it}(a)da$ of a cylindrical set of trajectories with the base $R\in \frak R^{(1)}$ (= the Borelian sets on the line) and the generatrix corresponding to piecewise constant function $\alpha_k (\tau )$, see (9).

\medskip

To construct the retarded Green's function $\fC_{it}(a_1 ,\ldots ,a_n )$ on n-dimensional base of the cylindrical set of trajectories (9) as a functional on $K$, consider the integral 
$$
\int{\(\prod\limits_{j = 1}^{n}{\ol{C_{i\Delta{t_j}}(\Delta{x_j})}}\)\vphi\(\sum\limits_{j = 1}^{n}{\alpha_{j1}\Delta{x_j}}, \ldots ,\sum\limits_{j = 1}^{n}{\alpha_{jn}\Delta{x_j}}\)d\Delta{x_1} \ldots d\Delta{x_n}} \text{,} \tag 14
$$
 where $\varphi (a_1 ,\ldots ,a_n)\in K$.

Repeating literally the construction of the   Green's function $\fC_{it}(a)$, see (12), we get the momentum presentation of the promised  Green's function 
$$
\wtfC_{it}(p_1 ,\ldots ,p_n) = \exp{\(i\sum\limits_{j = 1}^{n}{\abs{\sum\limits_{k = 1}^{n}{\alpha_{jk}p_k}}\Delta{t_j}}\)} = \exp{\(i\sum\limits_{j = 1}^{n}{\abs{(\alpha_{j},p)}\Delta{t_j}}\)}
\tag 15
$$
 as a  functional on $\psi (p_1 ,\ldots p_n )\in Z$.

\medskip

 In the r.h.s.~of (15) there is the product of functionals which are  Green's functions on one-dimensional base of the cylindrical set of trajectories. Thus
$ C_{it}(x)$ uniquely defines the measures of cylindrical set of trajectories
 in $R^{(n)}$ with $n$-dimensional Borelian bases.

 Here the complex measure of some cylindrical sets can have infinitely large imaginary part (of different signs), but the 
 measure of all cylindrical sets of translations with base of any given finite dimension is $\int\fC_{it}(a_1 ,\ldots ,a_n )da_1 \ldots da_n =1$ and the pre-measures are compatible, see \cite{3}.
 
 We will see (section 5, theorem (II)) that it is enough to consider in (11) only the matrices $\alpha_{jk}$ with $\Sigma_{j=1}^n \sqrt{\alpha^2_{j1}+\ldots +\alpha_{jn}^2} \Delta t_j$ bounded for every $n$, which implies (see (19) below, sections 5 and 6) that $\alpha_k (\tau )\in L_2 (0, t)$, see (11).
 
 Therefore one has 
$$
\ld\int f(a_k )\fC_{it} (a_k ) da_k \rv_{k\to\infty}= \int f(a )\fC_{it} (a ) da \tag 16
$$
 for every bounded continuous function $f(a)$ of $n$ variables. So the   measure of   cylindrical sets of trajectories defined by $
 \fC_{it} (a)$ satisfy the continuity condition (the weak continuity), see \cite{3}.

 \medskip

Therefore we have deduced 

{\bf Theorem (I)}: {\it The quantum Cauchy functional $C_{it}(x )$ on the bump functions  yields, as a measurable function, complex-valued normalized continuous  measures
of cylindrical sets of translations with finite-dimensional
  Borel bases (pre-measures), and the measures of different cylindrical sets are mutually compatible.}

\medskip

Let us show that there exists a topological vector space $X$ to whose dual space   the measure of cylindrical sets of trajectories from theorem 1 can be extended; let us construct such an $X$.

To that end we construct a set of infinite matrices $\alpha_{jk}$ that define $X$ (see (9)) such that the measure, corresponding to $\fC_{it}(a)$, 
 of the complement to the ball $\Sigma_{j=1}^n a_j^2 = R^2$ ($a_k =\Sigma_{j=1}^n \alpha_{jk}\Delta x_j$) and its every subset tends to zero for $R\to \infty$ for every $n$. This is possible (despite to the fact that  $\fC_{it}(a)$ is complex valued) since the corresponding measure of every subset of the above domain is purely negative imaginary (see below). 
 Therefore we will show that the promised topological space is the real Hilbert space to whose dual space $X'$ extends the pre-measure  of trajectories corresponding to  $\fC_{it}(a)$, i.e., the $\sigma$-additivity of that measure on that space.  

\medskip

One should point out that this approach to the problem of extension of measures is due to
 V.~D.~Erokhin (see \cite{13}, \cite{17}) who formulated a
 criterion for continuation of a probabilistic pre-measure on a countably-normed space $X$ to all its Borel subsets.

To estimate  the measure of the ball complements in the next section, we use an analytic construction due to R.~A.~Minlos (see \cite{13}).

\head 3. Estimating the quantum Cauchy functional measures of the complements of balls and their subsets \endhead

We find the measure of the complement to a ball of radius $R$ in $n$-dimensional space. Our measure  corresponds to $\fC_{it}(a_1 ,\ldots ,a_n )$ viewed as a measurable function on the Borel subsets of $R^{(n)}$; we use
its Radon transform $\Cis (\xi ;r)$, see \cite{18}. 

 At that
$\Cis (\xi ;r)$ (where $(\xi ,a)=r$, $\xi$ is a unit vector in $R^{(n)}$) is a measurable function   on $R^{(1)}$. 

Thus the problem is reduced to construction of a generalized function of single argument $(\xi ,a)$ that corresponds to $\Ci (a)$. 
To that end let us consider $\int \Ci (a) \varphi ((\xi,a ))da$, where $\varphi (r)\in K^{(1)}$. Recall that a similar problem was already solved when we constructed the  generalized function of argument $\Sigma_{j=1}^n \alpha_j \Delta x_j =a$ that corresponds to $\Ci (x)$.  Repeating the computation literally, we get 
$$
\Cis (\xi ;r)=C_{iQ}(r), \quad Q= \Sigma_{j=1}^n \abs{(\xi,\alpha_j )}\Delta t_j ,\quad (\xi,\alpha_j )=
\Sigma_{k=1}^n \xi_k \alpha_{jk} .\tag 17
$$
 Therefore 
$$
\Cis (\xi ;r)={1\over 2} (\delta (r+Q)+\delta (r-Q))+{i\over{\pi}}{Q\over{Q^2 - r^2}} .\tag 18
$$
 Thus the  measure  of the half-space $(a,\xi )\ge R >0$ of $n$-dimensional space, corresponding to $\Ci (a)$, equals $\int^\infty_R \Cis 
(\xi ;r)dr$.

But the measure of that half-space equals $\int\chi_R^\xi (a)\Ci (a)da$. where   $\chi_R^\xi (a)$ is the characteristic function of the half-space. So 
$$
\overline{\int_R^\infty  \Cis (\xi ;r)dr}^{S_1} =
 \overline{\int\chi_R^\xi (a)\Ci (a)da}^{S_1} =
\int  \overline{\chi_R^\xi (a)}^{S_1}\Ci (a)da  ,\tag 19
$$

where the overline means $\xi$-averaging over the surface of the unit sphere.

Notice that in $ \overline{\chi_R^\xi (a)}^{S_1}$ the vector $a$ is fixed, and we integrate over   vectors $\xi$ with $(a,\xi )\ge R$. These vectors form a cap on the sphere cut by the plane that has distance $R/\abs{a}$ from 0, so, using the Radon transform on the sphere (see \cite{18}), we see (cf.~\cite{3}, \cite{14})  that 
$$
 \overline{\chi_R^\xi (a)}^{S_1}= N_n \int^1_{R/\abs{a}} \(1-y^2 \)^{(n-1)/2} dy    \tag 20
$$
   if $ \abs{a}\ge R$ 
and 0 otherwise; here   constant $N_n$ depends on dimension $n$. Since $ \overline{\chi_0^\xi (a)}^{S_1}=1/2$, one has 
$$
N_n ={1\over 2}
\(\int^1_{0} \(1-y^2 \)^{(n-1)/2}dy\)^{-1} .   \tag 21
$$
  Using sperical coordinates in (19), we get

$$
\overline{\int_R^\infty\Cis (\xi ;r)dr}^{S_1} =N_n \int_R^\infty \int^1_{R/\rho} (1-y^2 )^{(n-1)/2} dy  d_{\rho} \int _{\abs{a}<\rho} \Ci (a)da .\tag 22
$$
 This equality relates mean values of measures   corresponding  to $\Ci (a)$ of half-spaces and of balls
 (compare with identical result   in \cite{3}, \cite{13}).

\medskip

Consider $ \overline{\int_R^\infty \Cis (\xi ;r)dr}^{S_1} $. For $R>Q$ (see (13))  the mean value theorem implies (recall that the integral is understood in the sense of the Cauchy principal value, see (13))  
$$
\overline{\int_R^\infty\Cis (\xi ;r)dr}^{S_1}  ={i\over\pi} \overline{
\int_R^\infty{Q\over{Q^2 - r^2}}dr}^{S_1} =-{{i}\over{2\pi}}\ln\abs{{{P+1}\over{P-1}}}, \tag 23
$$
 where $P={1\over R}\Sigma_{j=1}^n \abs{(\hat{\xi},a)}\Delta t_j >0$ and $\hat{\xi}$ is a fixed point on the unit sphere.

Therefore, using (22) and (23), we get

$$
-{{i}\over{2\pi}}\ln\abs{{{P+1}\over{P-1}}}=N_n \int_R^\infty \int^1_{R/\rho} (1-y^2 )^{(n-1)/2} dy  d_{\rho} \int _{\abs{a}<\rho} \Ci (a)da 
.\tag 24
$$

So the measure    of the complement to a ball of large enough radius, that corresponds to $\Ci (a)$, is negative purely imaginary for every $n$.

Notice that 
$$
N_n  \int^1_{R/\rho} \(1-y^2\)^{(n-1)/2} dy ={1\over 2} \( \int^{\sqrt{n}}_0 \(1-x^2\)^{(n-1)/2} dx\,\)^{-1} 
 \int^{\sqrt{n}}_{R\sqrt{n}/\rho} \(1-x^2\)^{(n-1)/2} dx.
$$
 So for $n$ large enough one has 

$$
N_n  \ld\int^1_{R/\rho} \(1-y^2\)^{(n-1)/2} dy\rv_{n\to\infty} \simeq 
{1\over 2} \( \int^{\infty}_0 \exp(-x^2 /2  ) dx\,\)^{-1} 
 \int^{\infty}_{R\sqrt{n}/\rho} \exp(-x^2 /2  )   dx =
$$
   
$$
 = \sqrt{1/2\pi}   \int^{\infty}_{R\sqrt{n}/\rho} \exp(-x^2 /2  )   dx  ,
$$

cf.~\cite{3}, \cite{13}.

Returning to (24), one has

$$
\ld{{1}\over{2\pi}}\ln\abs{{{P+1}\over{P-1}}}\rv_{n\to\infty}  \cong \sqrt{1/2\pi}   \int^{\infty}_{R } 
\int^{\infty}_{R\sqrt{n}/\rho } 
\exp(-x^2 /2  )   dx d_\rho \int_{\abs{a}<\rho} i\Ci (a)da ,
$$

and, since the inner integral in r.h.s.~decreases when $\rho $ decreases (and we deal with equality of positive numbers), we get

$$
\ld{{1}\over{2\pi}}\ln\abs{{{P+1}\over{P-1}}}\rv_{n\to\infty}  > \sqrt{1/2\pi}   \int^{\infty}_{\sqrt{n} } \exp(-x^2 /2  )   dx
\int_{\abs{a}\ge R}i\Ci (a)da .
$$
 This implies (see (17)) that 
$$
\lim_{R\to\infty}
\int_{\abs{a}\ge R}i\Ci (a)da=0 \tag 25
$$
 if $\alpha_{jk}$ are such that 
$$
\Sigma^n_{j=1}\abs{(\hat{\xi},\alpha_j )}\Delta t_j  \le \Sigma^n_{j=1}\sqrt{\alpha_{j1}^2 +\ldots + \alpha_{jn}^2 } \Delta t_j   <N \tag 26
$$
 for every $n$, see (23).

Using the above method, we study the local structure of the measure of the complement to that ball that corresponds to $\Ci (a)$, i.e., the measure of a neighborhood of any point $A$ that lies outside the ball of radius $R$. 

Let us find the measure of the ball of radius $\sigma$ with center at $A$ corresponding to $\Ci (a)$.  
To that end it is enough to know the Radon  transform $\Cis (\xi; r+(a,A))$ of the functional $\Ci (a+A)$ (see \cite{18}). Consider

$$
\Cis (\xi; r+(\xi ,A))={1\over 2} (\delta (Q-r-
(\xi ,A))+\delta (Q+r+
(\xi ,A)))+{i\over{\pi}}{Q\over{(Q^2 - (r+
(\xi ,A))^2}},
$$
 where $Q=\Sigma_{j=1}^n \abs{(\xi ,\alpha_j )}\Delta t_j$, see (12).

Repeating literally the previous arguments, we get

$$
\overline{\int_0^\sigma\Cis (\xi ;r+(\xi ,A))dr}^{S_1} =N_n \int_0^\sigma  \int^1_{0} \(1+y^2\)^{(n-1)/2} dy  d_{\rho} \int _{\abs{a}<\rho} \Ci (a+A)da . 
$$
 By the mean value theorem, this implies 
$$
 \overline{\int_0^\sigma\Cis (\xi ;r+(\xi ,A))dr}^{S_1} ={i\over{\pi}} \int_0^\sigma {{\hat{Q}}\over{\hat{Q}^2 -(
r+(\hat{\xi} ,A))^2}}dr, 
$$
 where $\hat{Q}=\Sigma_{j=1}^n \abs{(\hat{\xi} ,\alpha_j )}\Delta t_j$, and $
\hat{\xi} $ is a fixed point on the unit spere with center at $A$.
So for large enough translations $A$ (precisely, if $
\abs{r+(\hat{\xi} ,A)}>\hat{Q}$), and for an arbitrary small $\sigma >0$, the measure $\int_{\abs{a}<\rho<\sigma} \Ci (a+A)da$
of the ball we consider is purely imaginary.

\medskip
Therefore, if (26) holds, one has 
$$
\lim_{R\to \infty} \int_\zeta \Ci (a)da=0,\tag 27
$$

where $\zeta$ is any domain that does not intersect the ball of radius $R$ in $n$-dimensional space. 

By a similar argument  (26) also yields 
$$
\lim_{R\to \infty} \int_\zeta \ol{\fC}_{it} (a)da=0, 
$$

where $\int_\zeta \ol{\fC}_{it} (a)da$ is positive purely imaginary.

\head 4. The generalized quantum Cauchy measure of the Borel subsets of a Hilbert space  \endhead

Notice that (26) implies that for every $n$ the series $\Sigma_k \alpha_{jk}^2$ converges for each $j$ (i.e., $\alpha_{jk}\in l_2$ for each $j$). Thus for some
$A_j (\sigma )\in L_2 (0,s)$, $s\le t$, one has
 $\Sigma_{k=1}^\infty  \alpha_{jk}^2 =\int_0^s A_j^2 (\sigma )d\sigma$.
  Therefore condition (26) means that the integral $\int_0^t \sqrt{\int_0^s A_\tau^2 (\sigma )d\sigma }d\tau$ is defined, and since 
$$
\int_0^t \sqrt{\int_0^s A_\tau^2 (\sigma )d\sigma }d\tau \le \sqrt{t} \sqrt{\int_0^t  \int_0^s A_\tau^2 (\sigma )d\sigma d\tau}=
 \sqrt{t} \sqrt{\int_0^t  \int_0^s A^2 (\sigma ,\tau )d\sigma d\tau} , 
$$

this follows if $ A(\sigma ,\tau )$ is any element of the Hilbert space of functions on the rectangle. 

\medskip

Thus come the equalities 
$$
 \dot z (\sigma )=\int_0^t A  (\sigma ,\tau )d x (\tau )= \int_0^t A  (\sigma ,\tau )\dot x (\tau )d \tau , \tag 28 
$$
 which are functional analogs of (9).

This implies that  
$$
a = 
 \int_0^t \alpha  (\tau )\dot x (\tau )d \tau   
$$
  exists for every   $\alpha (\tau )\in L_2 (0,t)$; thus $\dot x (\tau )\in L_2 (0,t)$ (by the Riesz theorem about linear functionals on $L_2 (0,t)$, see \cite{16}).
 
Therefore, since $\alpha (\tau)$ form a real Hilbert space, the complex-valued cylindrical measures that correspond to $\fC_{it}(a_1,\ldots ,a_n )$ extend to its dual (which is the space of velocities $\dot x (\tau )\in L_2 (0,t)$). 

Indeed, the measure corresponding to $\fC_{it}(a)$ of a ball in the Hilbert space of large enough radius $R$ can be chosen arbitrarily close to 1 (see the previous section). Thus, since the ball is weakly compact (see \cite{3}), it can be covered by a finitely many cylindrical subsets. Hence, if for every infinite collection of nonintersecting cylindrical sets some of the covering them cylindrical sets contain infinitely many terms of that collection, the measure of their union equals the sum of their measures, since they have common generating subspace. Since up to an arbitrary small number the measure of the union of any collection of nonintersecting cylindrical sets equals the sum of their measures, this implies that the pre-measure corresponding to $\fC_{it}(a)$ extends to arbitrary Borel subsets of $L_2 (0,t)$ (not only to cylindrical subsets with finite-dimensional bases) and that the quantum Cauchy measure $\fC_{it}(a)$ is $\sigma$-additive.

 \medskip
 
Choosing appropriate basis $\alpha (s )\in L_2 (0,t)$ we find that the arguments $a$ of the complex-valued  $\fC_{it}(a)$ have physical interpretation as the values of coordinates on the trajectory at different moments of time 
$$
a(\tau) = \int_{0}^{t}{\alpha(s)\dot{x}(s)} = \int_{0}^{\tau}{dx(s)} = x(\tau) \text{,\quad} (0 \leqs \tau \leqs t) \text{.}
\tag 29
$$

Using those variables one can show that the measure that corresponds to the quantum Cauchy functional has  locally compact support in the uniform topology. Let us prove this.
 
 \medskip

 It is clear that every trajectory $x(\tau )=\int_0^\tau \dot x (s) ds$ from the support $[x_\tau ]$ is continuous. 
  Notice that for elements of that support one has (here we set  $\dot x (s)=0$ for $s\notin (0,t)$)
 
$$
\abs{x(\tau +T)-x(\tau )} =\abs{\int_0^{\tau+T} \dot x (s)ds -
 \int_0^{\tau} \dot x (s)ds}=
$$
 
$$
=\abs{\int_\tau^{\tau+T} \dot x (s)ds}
  \le \sqrt{T}\sqrt{\int_\tau^{\tau+T} {\dot x}^2 (s)ds} \le\sqrt{T}N,\quad N<\infty,
$$
 since $\dot x (\tau )$ are uniformly bounded in $L_2 (0,t)$, see (25).
  
  Therefore $\ld\sup_{0\le \tau \le t}\abs{x(\tau+T)-x(\tau )}\,\rv_{T\to 0} \to 0$ simultaneously for all $x(\tau )\in C(0,t)$ in the support.
 Thus the support $[x_\tau ]$ of measures of cylindrical subsets of $L_2 (0,t)$ that correspond to $C_{it}(x)$, is locally compact  in the uniform convergence topology of $C(0,t)$ by Arzela's theorem (see  \cite{16}).

{\bf Theorem (II)}: {\it The topological vector space on whose dual 
 the generalized pre-measure    that corresponds to the quantum Cauchy functional $C_{it} (x)$, is extendable as a measure of its arbitrary Borell sets, is a real Hilbert space; its dual is the space of velocities.
The support $[x_\tau ]$ of that  generalized complex $\sigma$-additive measure   is locally compact for the uniform convergence topology with $\dot x (t)\in L_2 (0,t)$.}

\medskip

Corollary. {\it The result automatically renders to the case of the space  quantum Cauchy functional.}

\head 5. The quantum generalized Cauchy functional and
the quantum generalized Cauchy process   \endhead

Consider the direct product of functionals in (15)

$$
  \int (\Pi_{j=1}^n \overline{C_{i\Delta t_j}(\Delta x_j )})\vphi ((A \Delta x)_1,\ldots ,(A\Delta x)_n )   d\Delta x_1 \ldots   d\Delta x_n  =   \tag 30
$$
   
$$
=   \int (\Pi_{j=1}^n \overline{C_{i\Delta t_j}(\Delta x_j )})\vphi ( (A\Delta x))   d  x_1 \ldots   d  x_n  .
$$
 Since the generalized measure that corresponds to $C_{it}(x)$ is $\sigma$-additive at $A \in L_2$ (theorem II), one can pass
 in (30) to the limit for $n\to \infty$ ($\max \Delta t_j \to 0$).
 We get (here $[x_\tau ]$ is the support of the measure) 
$$
\int_{[x_\tau  ]}(\Pi^t_{\tau =0} \overline{ C_{id\tau} (dx(\tau ))})\varphi (\ldots, dx_\tau,\ldots )\Pi^t_{\tau =0} dx_\tau , \tag 31
$$
 where $\varphi (\ldots ,dx(\tau ) ,\ldots ) =\varphi (\ldots ,\dot x (\tau )d\tau ,\ldots )\in K $ are bump functions of infinitely many variables (the bump functionals),   the argument of $\varphi$ (which is the functional variable of integration in (31))  is a trajectory $x(\tau )\in [x_\tau ]\subset C(0,t)$ since $dx (\tau )$ is an element of that trajectory.
 
Let us point out that   $dx(\tau )=\dot x (\tau )d\tau$ with $\dot x (\tau )\in L_2 (0,t)$, so $dx(\tau )$ are not independent, as opposed to $\Delta x_j$ in (30). 
There the detection of property of $\sigma$-additivity of the premeasures answering led to expansion of a finite set of the functional arguments of bump functions to a set of a measure support elements. 

\medskip

By (29) and theorem 2, we see tha (31) makes sense without the bump functional 
$\varphi (\ldots ,dx(\tau ) ,\ldots )$ and  $\int_{[ x_\tau ]}\Pi^t_{\tau =0} C_{id\tau } 
(dx(\tau )) dx_\tau  =1$.

Therefore we get the generalized complex quantum Cauchy measure $$\int_{W_\tau} \Pi^t_{\tau =0} C_{id\tau } 
(dx(\tau )) dx_\tau$$ of a Borel set of trajectories $W_\tau \subset    [ x_\tau ]$
 (the quantum Cauchy process). If $W_\tau$ is bounded in the uniform convergence topology then it is compact.

 Notice that the ``density of measure" $ \Pi^t_{\tau =0} C_{id\tau } 
(dx(\tau )) $ of a trajectory $x(\tau )\in [x_\tau ]$ makes no sense.

\medskip

Recall that we consider the evolution of a single quantum relativistic particle in Foldy-Wouthuysen sense. 
So, as a side remark, a ribbon-shaped $W_\tau$ can illustrate a track in Wilson camera if we don't take in account the influence of the vapor particles on the quantum particle.

We should also recall that the question about the possibility of description of quantum phenomena and their evolution in terms of trajectories was asked by R.Feynman (see \cite {9}), thus finding a way of alternative description of the evolution of a quantum particle not as a wave but as a point-like object moving along a collection of trajectories (the support) with complex weights. In the case considered above of a  quantum relativistic particle the Cauchy measure $\int_{W_\tau} \Pi^t_{\tau =0} C_{id\tau } (dx(\tau )) dx_\tau$ is a complex wight for each ribbon of trajectories $W_\tau$. 

{\it Digression}. It is important to point out also, that existence of generalized quantum Cauchy measure $\ld \(\Pi^t_{\tau =0} C_{id\tau }(dx(\tau ))\) \rv_{x(\tau) \in [x_{\tau}]}$ let us to expand ordinary concept of particle state in quantum physics as initial state in Cauchy problem for Schr\"odinger equation (in this case ${\pderby{t}}C_{it}(x) = -{{i} \over {\pi}} x^{-2} \mathop{*} C_{it}(x)$ -- see below). Namely, task of the functional $\ld \(\Pi^t_{\tau =0} C_{id\tau }(dx(\tau ))\) \rv_{x(\tau) \in [x_{\tau}]}$ on an interval of time $[0,t]$ will result unambiguous definition of this functional for all $T > t$. It is easily seen that it is instantaneously arise from satisfaction $\wtC_{it}(p) \cdot \wtC_{is}(p) = \wtC_{i(t + s)}(p)$ (see (15)), which at the same time is interpreted as realization of the principle of causality (in a finite-dimensional case).

Let's note that this arisen expanded concept of a condition of quantum system $\int_{W_\tau} \Pi^t_{\tau =0} C_{id\tau } (dx(\tau )) dx_\tau$

Let us also point out that $C_{it}(x_1 ,\ldots ,x_n )$ is the same  generalized functional (31) viewed on the set of elements of the support that take values $x_1 ,\ldots ,x_n$ at fixed moments of time $t_1 ,\ldots ,t_n$ is the many-time construct.

\medskip

Convolving all the factors in  (31) and using   (10) we get 
  
$$
\int_{[x_\tau  ]}(\Pi^t_{\tau =0} \overline{C_{id\tau} (dx(\tau ))})\varphi (\int_0^t dx_\tau  )\Pi^t_{\tau =0} dx_\tau
=\int \overline{C_{it} (x)}\varphi (x) dx   , \tag 32
$$
 or, in short, $C_{it}= 
\Pi^t_{\tau =0} *C_{id\tau}.$

This equality expresses  the fundamental solution of equation 
$$
{{\partial}\over{\partial t}} C_{it}(x)=- {i\over\pi} x^{-2}* C_{it}(x).
$$

\medskip

{\bf Theorem (III)}: {\it The quantum Cauchy  functional  $C_{it} (x)$, viewed as a measurable function, yields    the generalized complex normalized quantum Cauchy measure   $\int_{W_\tau}\Pi_{\tau=0}^t C_{id\tau} dx (\tau )dx_\tau  $ of every Borel subset $W_\tau \subset L_2 (0,t )$. The
  support 
 $[x_\tau ]$ of that measure is   locally compact in $C(0,t)$ with derivatives $\dot x (\tau)\in L_2 (0,t)$. }

\head 6. The modified quantum Cauchy functional and the modified generalized  quantum Cauchy measure  \endhead

The coordinate presentations of the quantum Cauchy  functional  $C_{it} (x)$
(5) and the modified quantum Cauchy  functional  $C^m_{it} (x)$ are related as follows. The ratio of their imaginary parts $im\sqrt{t^2 - x^2}K_1 (im\sqrt{t^2 - x^2})=B(t,x)$ is an infinitely differentiable function nowhere equal to  0 or $\infty$ and $\ld{B(t,x)}\rv_{\abs{x}=t} =1$. This implies  $\int\overline{C_{it}(x) B(t ,x)}\varphi (x)dx = \int\overline{C_{it}^m (x)} \varphi (x)dx $, i.e., 
$C^m_{it}(x)=C_{it}(x) B(t ,x)$ if $C_{it}(x) B(t ,x)$ is considered as a functional. One has $ B(t ,x)\varphi (x)\in K$, see \cite{2}, and $\ld{B(t,x)}\rv_{t\to 0} =1$ at the support $[x_\tau ]$ (due to its continuity), so the Fredholm denominator of the passage from $C_{it} (x)$ to $C_{it}^m (x)$ is finite. Therefore one has

\medskip

{\bf Theorem (IV)}: {\it The modified quantum Cauchy  functional  $C^m_{it} (x)$, viewed as a complex-valued measurable function, yields    the generalized complex modified quantum Cauchy measure   $\int_{W_\tau}\Pi_{\tau=0}^t C_{id\tau} dx (\tau )dx_\tau  $ of every Borel subset $W_\tau \subset L_2 (0,t )$. The
  supports of that measure and of the one  corresponding to the generalized  quantum Cauchy process are the same. 
  } 

 \medskip

 It is clear that Green's functions $ C^m_{it}(x)$ and 
$ \ol{C}^m_{it}(x)$ can be represented by the corresponding functional integrals (``path integrals", see (32)).

\head 6. The fundamental solutions of   Dirac equations for the free electron and of   Maxwell equations for the Einstein photon. \endhead

One gets from (2) the relation between the fundamental solution  $D^m_t (x)$ of the  Dirac electron equation in position coordinates and its Foldy-Wouthuysen presentation $D^{mF}_t (x)$ (see \cite{6}, \cite{2}) 
$$
D^m_t (x)=T^m (x)* D^{mF}_t (x) * T^m (x). \tag 33
$$
 Here $D^{mF}_t (x)$ yields a matrix-valued generalized quantum measure that has therefore the same properties as $C^m_{it}(x)$.

\medskip

Recall now that the retarded Green's function 
$D^m_t (x)$ viewed as a functional on $K$ can be obtained also from the retarded Green's function $G_t (x)$ of the Klein-Gordon equation (Pauli-Jordan formulas, see \cite{8}). Namely 
$$
D^m_t (x)=\gamma^0 (
\gamma^0 {{\partial}\over{\partial t}}   + (\gamma, \nabla )+im)G_t (x)
$$
  where $\gamma^0$ and $\gamma$ are  Dirac's $\gamma$-matrices (see \cite{14}) and

$$
G_t (x)={{\delta (l_t^2   )}\over{2\pi}} -{m\over{4\pi}}\theta (t-r){{J_1 ( l_t )}\over{l_t}};
$$
 here $\theta$ is the Heaviside step function and $J_1$ is the Bessel function. These formulas amount to formula (33): this can be easily seen by passing to the Fourier transforms. We also see that $D^m_t (x)$ is a finite functional, as opposed to $D^{mF}_t (x)$. At that 
$D^m_t (x)$ and $D^{mF}_t (x)$ are unitary equivalent.

\medskip

Therefore   formula (33) should be interpreted as a result of passage from the 
Green's function $D^{mF}_t (x)$ of the Dirac equation in the
Foldy-Wouthuysen variables,
that yields   a $\sigma$-additive measure with support $[x_\tau ]$, to
the 
Green's function $D^{m}_t (x)$ of the Dirac equation (1) that vanishes outside the 3-sphere $r=t$ (outside the light cone) which also 
yields   a $\sigma$-additive measure  with support $\{ x_\tau \}$ compact in $C(0,t)$ with derivatives $\dot x (\tau )\in L_2 (0,t)$.

Thus Dirac's electron, being a quantum object, while moving along the trajectories of the support and forming some of it state at moment $t$, though does not go outside the light cone, has arbitrarily large velocity $\dot x_\tau$ which is incompatible with being in Minkowski's world (see also the first summand in the Pauli-Jordan formula). At that the return to Minkowski's world occurs only in quasi-classical approximation (see the last section of the article).

\medskip

{\bf Theorem (V)}: {\it   The fundamental solution $
D^m_t (x)$ of the Dirac equation, viewed as a complex matrix-valued   generalized function, yields the   generalized quantum complex matrix-valued  Cauchy-Dirac measure    $\int_{W_\tau}\Pi^t_{\tau =0} D^m_{d\tau} (dx(\tau ))dx_\tau $ of the Borel subsets $W_\tau \subset L_2 (0,t)$
   with support $\{ x_\tau \}$  which is compact in $C(0,t)$ with derivatives in $L_2 (0,t)$. }

\medskip

It is known that the system of Maxwell's equations (6) for the photon field in Majorana variables has coordinate presentation of the fundamental solution as a functional on $K$ (see (8)) 
$$
M_t (x) = Q^+ (x)* M^F_t (x) * Q(x)\,\times\,
Q^+ (x)* \ol{M}^F_t (x) * Q(x), \tag 34
$$
 where $\times$ is the direct product of
$3\times 3$-matrices. Here $M^F_t (x)\times  \ol{M}^F_t (x)$ yields also a (diagonal)matrix-valued generalized quantum measure of the Borel  sets in 
$L_2 (0,t)$
with support $[ x_\tau ]$.
Hence, due to the unitary equivalence of 
the Foldy-Wouthuysen transform of solutions of the Maxwell equation and the usual solutions of these equations, 
the generalized matrix-valued measure of the Borel subsets of the Hilbert space that corresponds to 
  $M _t (x)$ has, similarly to as was discussed above in the case of the Dirac equation,    
     has support $\{ x_\tau \}$  which is compact in the uniform convergence topology
 (since the speed of light $c$ is  constant).  

Thus we see that it is impossible to describe single photons as objects of the Minkowski world if we describe their evolution by using trajectories. We get

\medskip

{\bf Theorem (VI)}: {\it The retarded Green's function $
M_t (x)$ of the Maxwell equation, viewed as a matrix-valued  complex generalized function, yields  the
generalized  matrix-valued  complex    quantum   Cauchy-Maxwell measure $\int_{W_\tau} \Pi^t_{\tau =0} \text{ M}^{m}_{d\tau} (dx(\tau )) dx_\tau    $ of the Borel sets $W_\tau \subset L_2 (0,t)$
  with the support $\{ x_\tau \}$   compact in $C(0,t)$ with derivatives $\dot x_\tau \in L_2 (0,t)$. }

\medskip

We will not write down our matrix-valued complex Green's functions $D^m_t (x)$ and M$_t (x)$ as functional integrals (``path integrals"); they have similar structure and properties to the presentation of Green's function $C_{it}(x)$ given in (32). 

\medskip

It is important to notice that, due to  the unitarity of the  operators $Q(x)$, $T(x)$, $T^m (x)$, the retarded Green's functions of Dirac's particles and photons have similar space-time microstructure determined by  the  quantum   Cauchy process regardless of the fact that these are fermions and bosons.

\medskip

Following Schr\"odinger, we point out a peculiarity  of the Dirac  electron movement (see \cite{5}, \cite{14}). Namely, one can measure exactly just one Cartesian component of the velocity, so one may doubt the existence of electron's space trajectories that we have discussed above.

In fact,   in the support of the generalized measure that corresponds to $D^m_t (x)$ one has $\dot x (\tau ) \in L_2 (0,t)$, so the velocity is well defined. But the velocity
of the electron at the support of the measure is not continuous, hence has no exact values at any moment of time (for the 1-dimensional case as well). Thus the 
peculiarity  of   Dirac's electron movement observed by Schr\"odinger is compatible with the structure of the set of trajectories $\{ x_\tau \}$ that electron takes during its evolution.

Therefore the spontaneous micro-movement of Dirac's electron with indeterminately high speed, called Zitterbewegung (see \cite{5}, \cite{14}) can be seen as a corollary of the movement along the trajectories in such a support.

{\it Digression}. Zitterbewegung (literally ``twitching movement"), an irregular movement of  Dirac's electron with speed faster than light, was first noticed by  Schr\"odinger \cite{5}.

We point out the general nature of that phenomenon: it occurs for  all relativistic quantum particles, both photons (bosons) and Dirac's particles (fermions), and is related to unified quantum micro-structure of space-time characterized by single support $\{ x_\tau \}$ that  governs their evolution regardless of mass or of spin.

Notice that the evolution of state of  relativistic quantum particles that occurs due to that micro-structure of space-time, is purely kinematic which amounts to the Dirac or Maxwell equation evolution. 

The found alternative   description of evolution of relativistic quantum particles
by   accounting the Borel sets of their trajectories (with complex weight) together with solving the corresponding wave equations supports in a deeper way   the wave-particle duality of de Broglie in quantum mechanics.

\medskip

As was pointed out in the introduction, when one returns from the 
 Foldy-Wouthuysen variables, the support of the measures of trajectories of quantum relativistic particles restricts tothe ball of radius $t$ in the space of continuous functions $C(0,t)$ while preserving the fundamental local property of that measure that the velocities lie in the Hilbert space.

It is important that this property of velocities on the support helps to preserve the usual understanding of the wave function of a quantum relativistic particle as a state in a fixed moment of time, thus discarding the principal
argument, formulated first in the works of L.D.Landau and R.Peierls \cite{19}, \cite{20} (see also \cite{1}, \cite{8}, \cite{14}), against the possibility of existence of non self-contradictory quantum theory of single relativistic particles, which would imply that that equations of relativistic quantum mechanics make sense only after the secondary quantization.

At that the fact that velocities at the support lie in the Hilbert space confirms that the considered 
relativistic quantum particles,  if their state is formed via the path integral, lie outside the Minkowski world, hence their relativistic quantum theory is not local.

Notice that the existence of entangled states of quantum particles and their experimentally found properties tell that such velocities of relativistic quantum particles at the formation of their state actually occur, so the deduced above fact of their belonging to the Hilbert space seems not to be an artifact.

The existence and properties of the support of the generalized $\sigma$-additive measures of Cauchy-Dirac and Cauchy-Maxwell corresponding to the fundamental solutions of the Dirac and Maxwell equations and constructed in the present article, can be seen as physically based reinterpretation of solutions of these equations that return us to understanding them as characteristics of the movement of single electrons and photons. And it is that very property of the support of the above   generalized quantum measures makes this reinterpretation possible.

It is important to point out that quantum particles at the support have the said velocity $\dot x (\tau )\in L_2 (0,t)$ at times $\tau \le t$, i.e., at the formation of the retarded Green's function at moment $t$ using path integral (the Green's functions themselves do not depend on the velocity $\dot x (\tau )$). And it is these formed Green's functions determine the solution of all traditional problems about the Dirac and Maxwell equations including those of secondary quantization.

Therefore the discovered possibility of reinterpretation of the Dirac and Maxwell equations as equations for single particle, connected to the possibility of construction of generalized quantum measures of Cauchy-Dirac and Cauchy-Maxwell, can be seen in the Dirac and Maxwell equations
themselves, it does not contradict the traditional approach to the solution of these equations, and can be considered simultaneously with it.

\head 7. The correspondence principle for the free Dirac electron \endhead

Consider the correspondence principle for the Dirac electron using the ``path integral"  description of the fundamental solution of the Dirac equation (cf.~(32)):

$$
 \int \overline{D^m_t (x)}\varphi (x)dx=\int_{\{ x_\tau \} }(\Pi_{\tau =0}^t \overline{D^m_{d\tau} (dx(\tau ) )})\varphi \(\int_0^t dx(\tau )\)\Pi^t_{\tau =0} d x_\tau , \tag 35
$$

or simply $D^m_t =\Pi_{\tau =0}^t * D^m_{d\tau}$.

It is easy to see  that (35) means that 
$$
 \int \overline{D^m_t (x)}\varphi (x)dx=\int \overline{T^m (\alpha ) D^{mF}_t (x) T^m (\beta )} \varphi (\alpha + x+\beta )d\alpha dxd\beta ,
$$
 or  $D^m_t =
T^m * D^{mF}* T^m$, where  
$$
\int \overline{D^{mF}_t (x)}\varphi (x)dx=\int_{\{ x_\tau \} }(\Pi_{\tau =0}^t \overline{D^{mF}_{d\tau} (dx(\tau) )})\varphi \(\int_0^t dx(\tau )\)\Pi^t_{\tau =0} d x_\tau , 
$$
 or   $D^{mF}_t =\Pi_{\tau =0}^t * D^{mF}_{d\tau}$.

Since $m=m_0 \hbar^{-1} $ (here $m_0$ is the invariant mass of the electron, see \cite{14}; we still assume that $c=1$),   one has $\lim_{\hbar\to 0} T^m (x)=\delta (x)$, see (2). Therefore $\ld{D^m_t}\rv_{\hbar\to 0} \simeq \ld{D^{mF}_t}\rv_{\hbar\to 0}$, so the quasi-classical approximation to a solution of the Dirac equation equals to  its quasi-classical approximation in the Foldy-Wouthuysen coordinates, which implies 
$$
 \tag 36
$$
 where (see (2), (3)) one has

$$
D^{mF}_t (x) =\pmatrix
C^m_{it}(x) &0&0&0\\
 0&C^m_{it}(x)&0&0\\
 0&0& {C}_{-it}(x)&0\\
 0&0&0& {C}_{-it}(x)
\endpmatrix 
$$

So, using the asymptotic presentation of the Macdonald function $K_1 (z)_{z\to\infty} \simeq \sqrt{{\pi}\over{2z}} \exp (-z)$ (see \cite{10}), we find that the 
 asymptotic presentation $\ld{C^m_{it}(x)}\rv_{\hbar\to 0}$ as a functional 
$$
\tag 37
$$
 (here, as above,  $l_t =\sqrt{t^2 - r^2}$, $f=\abs{x}$, $m_0 l_t$ is the eikanal, and $\varphi (x) \in K $) which have lost the singularity at the light cone due to the construction of the asymptotics. It is easy to see that the two equal components $D^{mF}_t (x)$ exponentially disappear  outside the light cone (and it is them who yield the quasi-classical limit).

Here $\int_0^t  m_0 dl_\tau$ is the action functional for the free relativistic particle on any trajectory $x_\tau \in \{ x_\tau \}$ inside the light cone.  
 So at $\hbar \to 0$, by the principle of stationary action (see \cite{9}), in the path integral (36) of the modified quantum Cauchy functional the   
 classical  trajectory is distinguished    among all other trajectories in the support.  Accordingly, the support of Green's function of the electron degenerates into a single trajectory of the free relativistic particle. At that Zitterbewegung disappears.

Notice also that in quantum relativistic case, as follows from (35), (37), the functional dependence of the averaged functional from the free particle action  is exponential only in quasi-classical approximation, as opposed to  non-relativistic situation (see \cite{9}). This is a discriminating trait of the  quantization of  relativistic particles.
 
\head Conclusion \endhead

The present work shows special role played by generalized quantum Cauchy processes not only for
 the analysis of solutions of the relativistic quantum mechanics equations, but also for understanding of the physical meaning of these equations,
  as well as, possibly, in formulation and solving of new problems there.

Using these processes one finds a fundamental connection between solutions
of the Maxwell and Dirac equations,  which, probably, can be related to  a possibility of construction of the F.~A.~Berezin  superinteraction theory (see \cite{4}). 

We also find that the time derivative of the support of the quantum Cauchy processes (i.e., the velocity of photon and electron on the trajectory) belongs to a Hilbert space. This observation, probably, can serve as a base for explanation of the known paradox of Einstein-Podolsky-Rosen, see \cite{12}, and leads to an adequate mathematical description of such facts as collapse of the wave function at the moment of measurement, entangled states, etc.

We explain a special role of these processes for the analysis of the passage from quantum relativistic problems to their classical relativistic and non-relativistic versions.

The author thanks R.~A.~Minlos for his  support 
and priceless conversations, as well as members of O.~G.~Smolyanov's seminar at the mathematical department of  Moscow State University for the  constructive and kind critique.

\end